# Spectral Domain Z-scan Technique

**XI ZENG,[1,#] PENGFEI QI,[1,2,#] PIN CHEN,[1] LIE LIN,[1] AND WEIWEI LIU[1,*]**

[1] *Institute of Modern Optics, Nankai University,Tianjin Key Laboratory of Micro-scale Optical Information Science and Technology, Tianjin 300350, China*
[2] *School of Physics, State Key Laboratory for Mesoscopic Physics, Academy for Advanced Interdisciplinary Studies, Collaborative Innovation Center of Quantum Matter, Nano-optoelectronics Frontier Center of Ministry of Education, Peking University, Beijing 100871, China*
[#] *Xi Zeng and Pengfei Qi contributed equally to this work*
*\* liuweiwei@nankai.edu.cn*

**Abstract:** Characterizing the nonlinear optical properties of various materials plays a prerequisite role in nonlinear optics. Among different methods, the well-known Z-scan technique and the modified versions have been recognized as a simple and accurate method for measuring both the real and imaginary parts of the nonlinear refractive index. However, all the Z-scan methods based on detecting small beam variations put forward a severe restriction on the roughness of materials. Therefore, measuring nonlinear optical properties of highly scattering media still remain challenging. Inspired by the innovation of conventional Z-scan method that converting the wavefront phase shift to the easily measurable spatial pattern in far-field, the alternative spectral domain Z-scan technique was presented in this paper. It has a great potential for highly scattering medium, based on the scattering efficiency is insensitive to the wavelength for Mie scattering as the wavelengths are far smaller than the roughness. Moreover, to demonstrate the advantages of spectral domain Z-scan technique, the nonlinear refraction of polished slides and frosted slides was measured, which agrees well with previous reports.

## 1. Introduction

Nonlinear optical (NLO) phenomena including second harmonic generation (SHG) [1], two-photon absorption (TPA) [2], stimulated Raman scattering [3], self-focusing (SF) [4-6], optical Kerr effect [7] etc, have attracted extensive interests in theory, experiment and application over last half century. The light behavior changed with the amplitude of the optical field during light-matter interaction is manifested as the essential feature in the nonlinear (NL) optics field, which presents as the light-dependent change of the parameters such as the absorption coefficient, refractive index, and light scattering properties of materials. It means that the NL refractive index and absorption coefficient are fundamental parameters in all the field of NL optics. As a consequence, the measurements of these NLO parameters play a prerequisite role in the investigation of NLO phenomena. Moreover, the breakdown induced by femtosecond laser have provided a highly precise and secure cutting with minimal collateral damage for surgery, as well as the NLO effects including coherent anti-Stokes Raman scattering [8], second-harmonic generation [9] and two-photon fluorescence [10] have been demonstrated a bright prospect to biological imaging and diagnostic. Extracting nonlinearities in highly scattering media such as biological samples has attracted great attention once again. Hence, the present work is devoted to characterize the nonlinear refractive index of high scattering media induced by the optical Kerr effect.

   In the past decades, various techniques including elliptical polarization method [11], nonlinear interferometry [12], multi-wave mixing method [13], beam distortion method [14] etc, to characterize NLO materials, yielding direct information on the nonlinearities and its origin have been presented. Among the above methods, the well-known Z-scan technique is a simple and accurate method for measuring both the real and imaginary parts of the NL refractive index of optical materials, which has been widely employed to characterize the NL properties of a wide range of optical materials [15, 16]. Many modified versions have been exploited since the presentation of the original Z-scan technique by Sheik-Bahae *et al*. [17]. Specifically speaking, the top-hat beam Z-scan [18], eclipsing Z-scan [19] and off-axis Z-scan [20] were introduced to improve the sensitivity; the reflection Z-scan technique was designed for high-absorbing material [21]; the two-color Z-scan method was presented to characterize the frequency nondegenerate nonlinearity [22]; the dual-beam, single-beam time-resolved Z-scan and thermally managed Z-scan [23-25] were developed to observe the temporal evolution of optical nonlinearities [26, 27].

However, all the above methods based on detecting small beam variations put forward a severe restriction on the roughness of materials. Therefore, highly scattering media, such as biological tissue (other than clear liquids such as vitreous humor) are not suitable for studies with this technique. For example, a doubtful conclusion that the corneal tissue (typically contains ~78% water) NL refractive index ($2\times10^{-19}$ m$^2$/W) is about one order of magnitude larger than the one of water ($2\times10^{-20}$ m$^2$/W [28]) has been obtained from the measured imperfect Z–scan curve arising from the surface imperfections. Therefore, many aggressive efforts have been devoted to develop a modified Z-scan to characterize the NLO properties of rough media. In 1990, Sheik-Bahae *et al* have demonstrated that the impacts of surface imperfection can be reduced by subtracting a low irradiance background Z-scan from the high irradiance scan, because that the undesired effects from surface imperfection are not dependent on the intensity of irradiance [15]. However, such method is invalid for the high scattering media, that the wavefront distortion induced by surface imperfection is much larger than the one induced by self-phase modulation (SPM). In 2015, a scattered light imaging method (SLIM) was applied to measure the NL refractive index of scattering media, based on the analysis of the divergence angle varies induced by self-focusing effect through the side-view images of the laser beam propagating inside highly scattering liquid suspensions. [29] However, the sensitivity of SLIM (the NL refractive index of ~$10^{-19}$ m$^2$/W) will be limited by the small change in divergence angle through the side-view images at close range. It might not be suitable for the inhomogeneous scattering medium and the widespread medium with the NL refractive index <$10^{-19}$ m$^2$/W. Moreover, the spectral re-shaping technique has been also developed to measure NL refractive index and NL absorption in a highly scattering environment with good sensitivity, utilizing the change in frequency domain induced by SPM, rather than in the spatial domain as in Z-scan [30].

The design concept of conventional Z-scan method is that converting the wavefront phase shift induced by light-matter interaction to the easily measurable spatial pattern in far-field, [31] which can be also called as spatial domain Z-scan technique. Similarly, considering the change in spectral domain induced by wavefront phase shift, the more straightforward spectral domain Z-scan technique is proposed, which has a great potential for highly scattering medium based on the scattering efficiency is insensitive to the wavelength for Mie scattering as the wavelength are far smaller than the roughness. This point has been demonstrated by the spectral re-shaping technique [30]. This modified Z-scan has a great prospect to characterize the NLO properties of biological samples and unpolished advanced material.

## 2. Theoretical model

In order to quantitatively describe the physical scenario of pulse-matter interaction, we can assume that a Gaussian pulse propagating in a NLO medium along $+z$ direction, the optical pulse can be written as:

$$E(z,t) = A(z)\exp(-t^2/2\tau_0^2)\exp\left[i(k_0 z - \omega_0 t)\right] \qquad (1)$$

Considering the spatial distribution of Gaussian pulse is a TEM$_{00}$ Gaussian beam of waist radius $w_0$, $A(z)$ can be written as $A_0 w_0/w(z)$, where $w(z)=w_0(1+z^2/z_r^2)^{1/2}$ and $z_r=kw_0^2/2$ is the diffraction length of the beam, $k=2\pi/\lambda$ is the wave vector, and $\lambda$ is the laser wavelength, all in free space. As is well known, providing the medium is "thin" enough, whose criterion is medium thickness $L<z_r$, the beam diameter change within the medium induced by both diffraction and NL refraction effects can be neglected. The optical intensity $I_0$ and phase change $\Delta\phi$ of the electric field as a function of $z'$ (the propagation depth in the medium) are now governed in the slowly varying envelope approximation (SVEA) by a pair of simple equations [15, 16]:

$$\frac{d\Delta\phi}{dz'} = k\Delta n(I) \tag{2a}$$

$$\frac{dI}{dz'} = -\alpha I \tag{2b}$$

For simplicity, only the Kerr nonlinearity was considered and the refractive index change $\Delta n=n_2 I$, thus phase change $\Delta\phi$ can be obtained as following:

$$\begin{aligned}\Delta\phi &= kn_2 I_0 \left(1+z^2/z_r^2\right)^{-1} \exp(-t^2/\tau_0^2)\int_0^L e^{-\alpha z'}dz' \\ &= \left(1+z^2/z_r^2\right)^{-1}\exp(-t^2/\tau_0^2)\Delta\phi_{max}\end{aligned} \tag{3}$$

Where $I_0=A_0^2$ is the peak intensity at the focus, $\Delta\phi_{max}$ is the maximum phase shift along propagation direction, that is, the phase shift induced by the peak intensity at the focus, which can be written as

$$\Delta\phi_{max} = kn_2 I_0 L_{eff} \tag{4}$$

Where $L_{eff}=(1-e^{-\alpha L})/\alpha$, with $L$ the medium thickness and $\alpha$ the linear extinction coefficient arising from absorption and scattering. Therefore, the output optical field from medium contain the NL phase distortion induced by SPM can be written as

$$E_{out}(z,t) = E(z,t)\exp(-\alpha L/2)\exp(-i\Delta\phi) \tag{5}$$

From a formal point of view, the frequency spectrum after self-phase-modulation can be obtained by the Fourier transform of the pulse amplitude[32]

$$F(\omega) = \frac{1}{2\pi}\int_{-\infty}^{\infty} E_{out}(t)\exp(-i\omega t)dt \tag{6}$$

It is well known that converting the NL wavefront phase shift to the easily measurable spatial pattern of far-field is the core concept of conventional Z-scan method[31]. Similar to the so-called spatial domain Z-scan technique, we can naturally design a spectral domain Z-scan technique based on the SPM induced spectral changes, as shown in Eqs. (5) and (6). It is more suitable for highly scattering medium on account of the scattering efficiency is insensitive to the wavelength for Mie scattering as the wavelengths are far smaller than the roughness.

Subsequently, the schematic diagram of set-up for spectral domain Z-scan technique is designed, as shown in Fig. 1. The beam is divided into two beams by a beam splitter, the weak transmission energy is focused on a power meter to monitor input power and the strong reflective beam is focused by a lens with a focal length $f$. The sample is moved through the focus by a motorized precision translation stage. Then the exiting light field containing the NL phase distortion induced by SPM is focused on the integration sphere and then coupled in a spectrometer. It is worth mentioning that by using the integration sphere, the fluctuation of the outcome of the spectrometer was reduced significantly, providing good sensitivity for evaluating the weak spectral change. As compared with the work introduced by Ref. [30], the experimental setup shown in Fig. 1 is simplified without using a delicate spectral re-shaping system.

The transmitted power through the filter can be obtained by the spectral integral of $F(\omega)$ in the bandpass $\lambda_1<\lambda<\lambda_2$, and the normalized spectral domain Z-scan transmittance $T(z)$ can be calculated as

$$T(z) = \int_{\lambda_1}^{\lambda_2} F(\lambda, Z)d\lambda \bigg/ \int_{\lambda_1}^{\lambda_2} F(\lambda, Z\to\infty)d\lambda \quad (7)$$

Where $F(\lambda, z)$ is the exiting spectrum when the sample located at $z$, and $F(\lambda, z\to\infty)$ is equivalent to the incident spectrum. For the spectral domain 'aperture' Z-scan (like the spatial domain aperture Z-scan), the $\lambda_1$ and $\lambda_2$ are the selected wavelength.

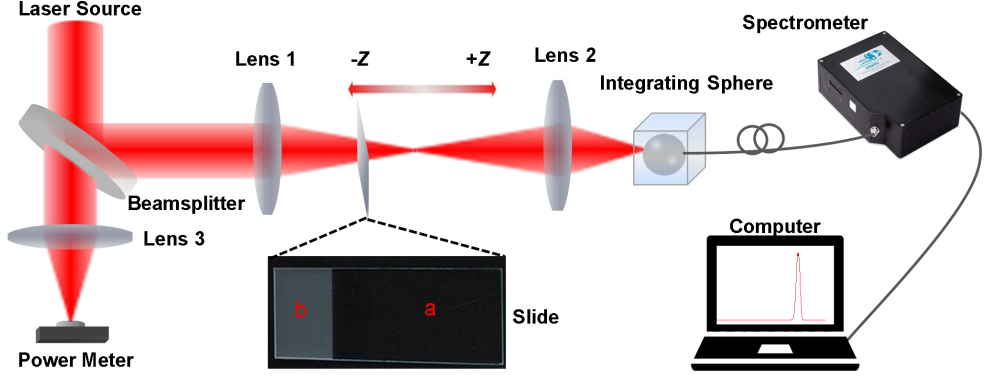

**Fig. 1.** The schematic diagram of set-up for spectral domain Z-scan technique.

### 3. Results and discussion

Now we will quantitatively discuss the spectral domain Z-scan technique. Considering the spectral width and the induced SPM, the femtosecond laser is an optimal choice for the spectral domain Z-scan technique. In our simulation, the center wavelength, pulse duration and pulse energy are 801.5 nm, 50 fs and 100 nJ, respectively. As shown in Fig.1, The beam radius before lens 1 is set as 2.5 mm, and the focal length f of lens 1 is set as 200 mm. The diffraction length of the beam after lens 1 is about 1.6 mm and the sample thickness is 1 mm.

Based on Eqs. (5) and (6), the exiting spectrum correspond to different $\Delta\phi_{max}$ can be easily obtained, as shown in Fig.2(a), which correspond to the output spectrum at position z=0 in spectral domain z-scan. Then based on Equation (7), the normalized spectral domain Z-scan transmission in the band of 800 nm ~ 803 nm are shown in Figure 2 (b). Similar to spatial domain aperture Z-scan, the linear transmittance of filter $S$ can be defined as:

$$S = \int_{\lambda_1}^{\lambda_2} F(\lambda, Z\to\infty)d\lambda \bigg/ \oint F(\lambda, Z\to\infty)d\lambda \quad (8)$$

where $\oint F(\lambda, Z\to\infty)$ represents the total spectral energy of entire spectral bandwidth. It can be calculated that the linear transmittance for the band of 800 nm to 803 nm is $\log_{10}(S) = -1.09$.

For $\log_{10}(S) = -1.09$, the Z-scan transmittance curves of different $\Delta\phi_{max}$ can be obtained theoretically, as shown in Fig. 2(b). The normalized transmittances $T(z)$ presents as single valley at focus. It should be noted that $T(z)$ presents as a peak and valley with a certain distance for spatial domain Z-scan, where the spatial diffractive pattern of far-field is determined by both lens effects induced by SPM and the distance between induction lens and focus.

The relationship among the linear transmittance of filter $S$, the maximum phase shift $\Delta\phi_{max}$ and the amplitude valley $T_V$ is very critical to extract NL refractive index $n_2$ according to Eq. (4). The $T_V$ defined as

$$T_V = 1 - Min(T) \tag{9}$$

The dependence of the amplitude of valley $T_V$ on the maximum phase shift $\Delta\phi_{max}$ under the specific linear transmittance of filter $S$ ($\log_{10}(S)$= −0.13, −0.26, −0.79, −0.86, −1.1, −1.2 and −1.9) are depicted in Fig.2(c). The overlapped curves demonstrate that the dependence of the amplitude of $T_V$ on the maximum phase shift $\Delta\phi_{max}$ is insensitive to the linear transmittance for $\log_{10}(S) = -0.79 \sim -1.9$. It should be noted that the use of Z-scan method often adopt low energy pulses to avoid both material damage and appearance of high-order nonlinearities. Considering the 'thin' sample ($L<z_r$), the maximum phase shift $\Delta\phi_{max}$ is generally less than $\pi$ for Z-scan, so that the range of $0<\Delta\phi_{max}<2.75$ is considered in this paper. The common used range of $S$ is 1%~10%, that is $-2<\log_{10}(S)<-1$.

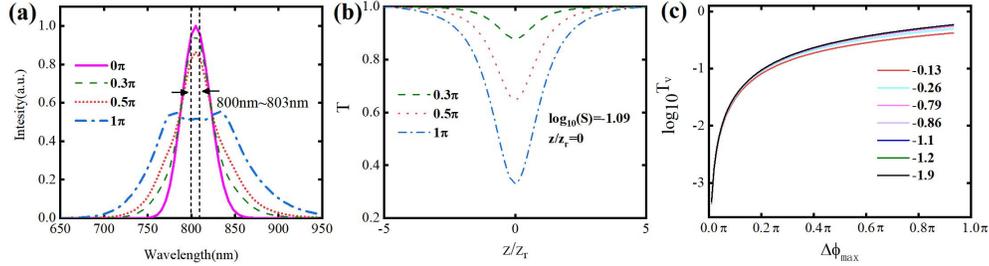

Fig. 2. Simulation results of spectral domain Z-scan technology. (a) The exiting spectrum correspond to $\Delta\phi_{max}$ =0, 0.3π, 0.5π and π. (b) The calculated Z-scan transmittance curves correspond to $\Delta\phi_{max}$ = 0.3π, 0.5π and π for $\log_{10}(S)$= − 1.09. (c) The dependence of the amplitude of $T_V$ on the maximum phase shift $\Delta\phi_{max}$ under different linear transmittance of filters, $\log_{10}(S)$ = −0.13, −0.26, −0.79, −0.86, −1.1, −1.2 and −1.9.

Subsequently, to demonstrate the advantages of our spectral domain Z-scan technique, the nonlinearity of the slide was measured. The sample used in the measurement is a fused silica based slide (sailboat 7105). One side of the slide was a polished glass and the other was a frosted glass, where a located at the polished region, and b located at frosted region, as shown in Fig. 1. Moreover, the energy loss when the light transmits through the frosted part was characterized to be about 23 %. The beam radius before lens 1 was set as 2.5 mm, and the focal length f of lens 1 was set as 200 mm. A Ti: sapphire femtosecond laser amplifier (Spitfire, Spectra-physics) was employed to generate 50 fs, 801.5 nm, 0.65 mJ, 1 kHz laser pulses. The focal length of lens 1 was 200 mm, and the beam waist $w_0$ at the focus was about 20.1 μm. Due to the small numerical aperture of the optical fiber and the spatial dispersion of scattered light, the exited spectrum was firstly focused into the fiber optic integrating sphere (FOIS-1, Ocean optics) by the lens 2, and then coupled in spectrometer (HR-2000, Ocean Optics). The sample was moved near the focal point with a step size of 0.1 mm, and the average of the three spectral signals was used as the experimental result for each Z position. Because the scattering and absorption of the sample reduces the spectral energy, the collected spectrum was normalized to guarantee the total spectral energy is 1, according to conservation of energy.

Firstly, the position *a* that located at the polished region was measured. Figs. 3(a1)-3(a3) depict the normalized spectral domain Z-scan transmittance $T(z)$ under the three different input average power. The incident average powers onto the samples were 67 μW (single pulse energy of 67 nJ), 75μW (single pulse energy of 75 nJ), and 140 μW (single pulse energy of 140 nJ), the corresponding peak intensities at the focus are about 210 GW/cm², 246 GW/cm² and 440 GW/cm², respectively. The corresponding experimental data can be well fitted by red lines. To visually present the SPM induced spectrum change in our measurements, the normalized spectrum recorded by the spectrometer for the sample locates at $z/z_r = -4.5$ and $z/z_r = 0$ are shown in Figs. 3(b1)-3(b3), where the selected band of 800 nm

~ 803 nm was labelled by blue dotted lines. The Phase shift $\Delta\phi_{max}$ and $n_2$ at three different power can be obtained by the transmission $T(z)$ measured in Fig.3(a), based on the theoretically calculated relationship between $\Delta\phi_{max}$ and the amplitude valley $T_V$ (Fig.2(c)). Therefore, we can get the $\Delta\phi_{max} = 0.439$ and $n_2 \approx 2.56 \pm 0.04 \times 10^{-16} cm^2/W$ for the peak intensity of 210 $GW/cm^2$, $\Delta\phi_{max} = 0.49$ and $n_2 \approx 2.75 \pm 0.12 \times 10^{-16} cm^2/W$ for the peak intensity of 246 $GW/cm^2$, $\Delta\phi_{max}=0.96$ and $n_2 \approx 2.79 \pm 0.04 \times 10^{-16} cm^2/W$ for the peak intensity of 440 $GW/cm^2$. The Nonlinear refractive index coefficients $n_2$ obtained at three intensities are perfectly consistent with each other.

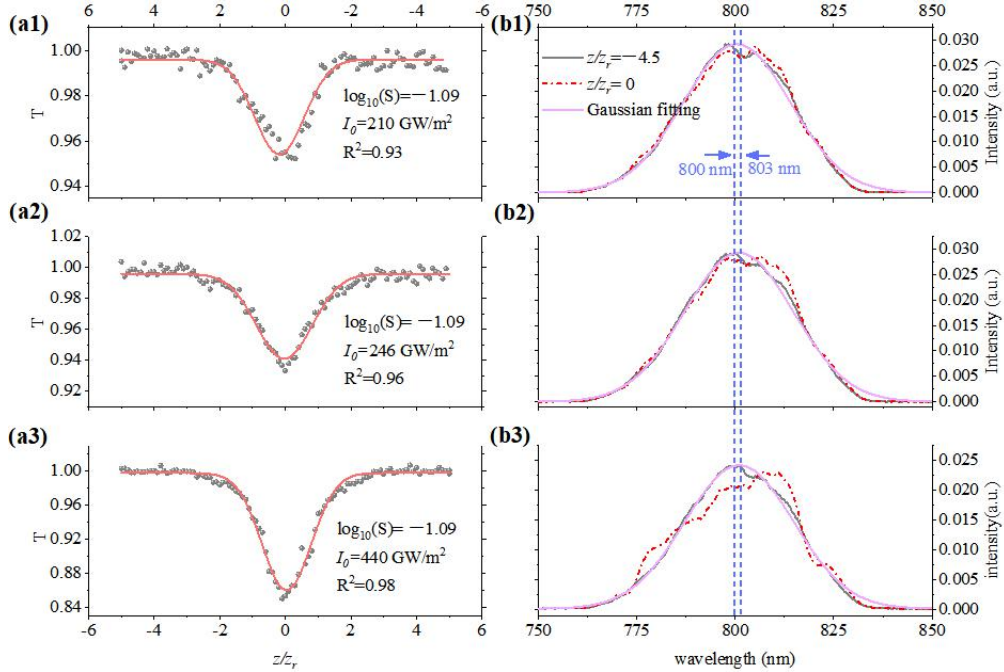

**Fig. 3.** (a) The normalized spectral domain Z-scan transmittance $T(z)$ of polished position $a$ under the three different peak intensities, i.e., $I_0 \approx 210$ $GW/cm^2$, $I_0 \approx 246$ $GW/cm^2$ and $I_0 \approx 440$ $GW/cm^2$. The red lines are the fitting curves; (b) the corresponding normalized spectrum recorded by the spectrometer for the sample locates at $z/z_r = -4.5$ and $z/z_r = 0$. The selected band of 800 nm ~ 803 nm was labelled by blue dotted lines

In order to demonstrate the advantage of our method for measuring the nonlinear refractive index of the scattering medium, the position b located at the frosted region of slide was then measured. Similarly, the nonlinearity refractive index at three different pulse energy was measured. Since the scattering of the frosted glass, the directional laser field was dispersed. In order to collect all spectral components with the same efficiency, the exiting light was first focused and coupled into the integrating sphere by lens 2. Then the spectrum modulated by SPM was recorded by the spectrometer. As shown in Fig. 4, the experimental data in fact are more fluctuated around the focal position. Frankly speaking, the exact reason is not clear yet. It might be due to the change of the beam size/roughness ratio during the z-scan. Around the focal point, the beam size/roughness ratio is small, affecting the precise value of the energy loss associated with the unhomogeneous roughness in micro scale. This effect might be particularly pronounced if the translation stage was not perfectly parallel to the laser beam propagation direction. In this sense, a fitting may be helpful in reducing the side effect of this kind of data fluctuation. Figs. 4(a1)-4(a3) shows the normalized spectral domain Z-scan transmittance $T(z)$ under the three different peak intensities, i.e., 203 $GW/cm^2$,

250 GW/cm² and 290 GW/cm². The normalized transmittance curves of the frosted glass slide in Figs. 4(a1)-4(a3) clearly present valley-shaped distribution. Although the signal-to-noise ratios of the normalized transmittance curves become worse than that of a position (Figs. 3(a1)-3(a3)), the experimental data still can be fitted by the red curves. It implies that the spectrum change induced by optical nonlinearity wasn't overflowed by the light scattering. According to the fitted curve of transmittance at three different peak intensities, the nonlinear refractive index coefficients $n_2$ can be obtained as $2.94 \pm 0.24 \times 10^{-16}$ cm²/W, $2.72 \pm 0.18 \times 10^{-16}$ cm²/W and $2.66 \pm 0.23 \times 10^{-16}$ cm²/W, respectively. The nonlinear refractive index coefficients of the positions a and b are similar, the average value of $\approx 2.73 \times 10^{-16}$ cm²/W also is in close agreement with three-wave frequency mixing measurement of $2.98 \times 10^{-16}$ cm²/W as reported in Ref. [33]. It prove that the spectral domain Z-scan method can be used to measure the nonlinear refractive index coefficient of the scattering medium.

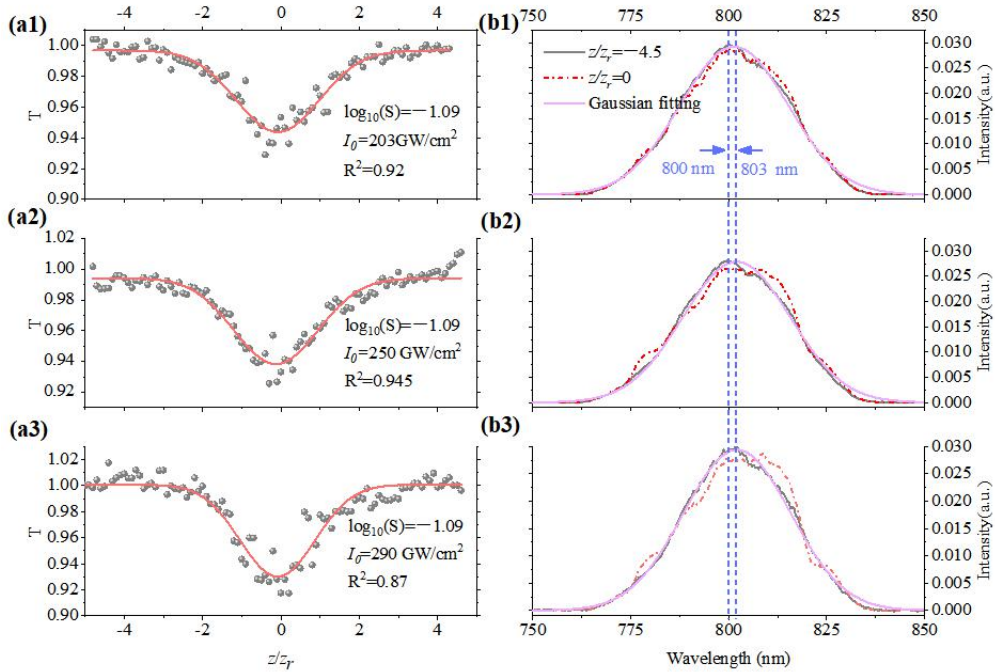

Fig. 4. (a) The normalized spectral domain Z-scan transmittance $T(z)$ of frosted position $b$ under the three different peak intensities, i.e., $I_0 \approx 203$ GW/cm², $I_0 \approx 250$ GW/cm² and $I_0 \approx 290$ GW/cm². The red lines are the fitting curves; (b) the corresponding normalized spectrum recorded by the spectrometer for the sample locates at $z/z_r = -4.5$ and $z/z_r = 0$. The selected band of 800 nm ~ 803 nm was labelled by blue dotted lines

## 4. Discussion

### 4.1 How to choose spectral window

Similar to the so-called spatial domain Z-scan technique, we designed a spectral domain Z-scan technique based on the SPM induced spectral changes. As is well known that the spatial domain Z-scan can be divided into conventional Z-scan and eclipsing Z-scan. The sensitivity was greatly enhanced in eclipsing Z-scan by choosing the edge region to monitor the change of spatial pattern in far-field induced by self-focusing or self-defocusing, whereas the center region is chosen in conventional Z-scan. As a consequence, choosing the spectral window is also an important issue for the spectral domain Z-scan. The spectral domain Z-scan can combine with eclipsing method like spatial domain eclipsing Z-scan by selecting spectra edge as spectral window, as shown in Fig. 5(a). Based on the calculated exiting spectrum, the normalized spectral domain eclipsing Z-scan transmittance $T(z)$ can be obtained, as shown in Fig. 5(a), which presents as a peak and the $T_p$ defined as

$$T_p = Max(T) - 1 \tag{10}$$

Then the dependence of the amplitude of $T_p$ and $T_v$ on the maximum phase shift $\Delta\phi_{max}$ and the linear transmittance $S$ for spectral domain eclipsing Z-scan and spectral domain Z-scan were calculated and depicted in Figs. 5(a) and 5(b), respectively. The label of color bar coordinate is $\log_{10}(T_p)$ and $\log_{10}(T_v)$. It can be clearly seen that the spectra change $T_p$ of edge region is much larger than center region under the same condition, that is, the sensitivity of edge region is much larger than center region, which is consistent with the situation of spatial domain Z-scan. However, considering the intensity of edge region is much weaker than center region, it is so difficult to extract the nonlinear signal from noises, thus the center region was chosen as spectral window in this work.

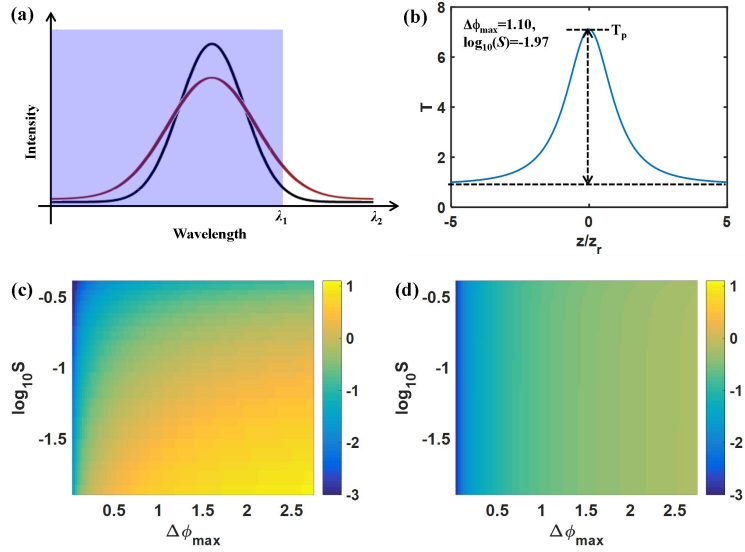

**Fig. 5.** (a) The schematic diagram of spectral domain eclipsing; (b) The calculated Z-scan transmittance curves for spectral domain eclipsing Z-scan; (c) and (d) The dependence of the amplitude peak or valley $T_p$ and $T_v$ on the maximum phase shift $\Delta\phi_{max}$ and the linear transmittance of filter $S$ for (c) spectral domain eclipsing Z-scan and (d) spectral domain Z-scan. The label of color bar coordinate is $\log_{10}(T_p)$ and $\log_{10}(T_v)$.

### 4.2 Pulse shape and high order nonlinearity

It should be noted that the theoretical model in this work was established by assuming the pure Gaussian pulse incident on the thin sample, and the input power is low enough that only the third order nonlinearity need to be considered. As is well known that the ideal Gaussian femtosecond laser pulse is very difficult to obtain. In our experiment, the input pulse can be well fitted by Gaussian function and the R Square is larger than 0.99, as shown in the blue dotted lines in Figs 3(b) and 4(b). The pulse shape of our laser system in temporal domain has been confirmed as Gaussian with pulse duration of 50 fs by the autocorrelator measurement. It means that the input pulse can be approximated as Gaussian pulse. Therefore, the third order nonlinear index coefficient $n_2$ can be rather accurately calculated from the experimental data, based on the simplified theoretical model described in section 2.

### 4.3 Two-photon absorption

It is worth noting that two-photon absorption (TPA) was not discussed in our current analysis. Inspired by M. Sheik-Bahae, et. al.'s early work in 1990 [15], one possible way to retrieve the information of TPA could be to compare the spectral domain z-scan measurements at two different input laser power similar to Fig. 3(a). The evolution of *T* at low laser power as function of the scanning distance will be very similar to the case when there is no TPA effect. However, since the energy loss during the transmission through the sample will depend on the laser intensity, the curve will be more distorted at higher laser power, with lower value of *T* at the focal point. Hence, by comparing the measurements, one may get the information of TPA. Moreover, the two-photon absorption coefficient can be also obtained by taking the total wavelength as spectral window for our method, similar to the open-aperture scan mode of spatial domain Z-scan.

## 5. Conclusion

In conclusion, the alternative spectral domain Z-scan technique was presented in this paper, inspired by the innovation of conventional spatial domain Z-scan technique that converting the wavefront phase shift to the easily measurable spatial pattern in far-field. The advantages of our spectral domain Z-scan technique were demonstrated experimentally. Based on the insensitivity of the scattering efficiency to the wavelength for Mie scattering, the spectral domain Z-scan technique has a great prospect to solve highly scattering problems and pave the way to characterize the NLO properties of biological samples and unpolished advanced material. Furthermore, similar technique in principle could be applied to the long pulse case as long as the performances, such as the resolution and the stability, of the spectrometer could fulfilled.

**Funding**

National Key R&D Program of China (2018YFB0504400), Tianjin Research Program of Application Foundation and Advanced Technology of China (19JCYBJC16800), Fundamental Research Funds for the Central Universities, Tianjin Special Program for Talent Development, Open Research Funds of the State Key Laboratory of High Field Laser Physics, Shanghai Institute of Optics and Fine Mechanics (SIOM), 111 Project (B16027) and Project funded by China Postdoctoral Science Foundation (2019M660283).

**Disclosures**

The authors declare no conflicts of interest.